\documentclass[a4paper]{article}
\usepackage[left=2.7cm,
			right=2.7cm,
			top=2.05cm,
			paperwidth=210mm, 	 		  
			paperheight=297mm]{geometry}

\usepackage{hyperref}
\usepackage[T1]{fontenc} % if needed
\usepackage[utf8]{inputenc} 
\usepackage{amsmath}         %writing math formulas
\usepackage{amssymb}%superset of the amsfont package, set of msam and msbm fonts
\usepackage{amsthm}
\usepackage{bbm}             %fuer bold math
\usepackage{bm}              %fuer bold math
\usepackage[dvips]{graphicx}
\usepackage[usenames]{color} %Here the option [usenames] causes the definitions
\usepackage[mathscr]{eucal}  %use of Euler script, Gothic letters with \mathscr
\usepackage{blindtext}
\usepackage{appendix}

\usepackage[square,numbers]{natbib}
\usepackage{authblk}

\title{Improving center vortex detection by usage of center regions as guidance for the direct maximal center gauge}

\author[1]{Manfried Faber}
\author[1]{Rudolf Golubich}

\affil[1]{Atominstitut, Techn. Univ. Wien, Austria}

\date{}                     %% if you don't need date to appear
\begin{document}
\maketitle

\begin{abstract}
The \textit{center vortex model} of quantum chromodynamic states that vortices, closed color-magnetic flux, percolate the vacuum. Vortices are seen as the relevant excitations of the vacuum, causing confinement and dynamical chiral symmetry breaking.  In an appropriate gauge, as \textit{direct maximal center gauge}, vortices are detected by projecting onto the center degrees of freedom. Such gauges suffer from Gribov copy problems: different local maxima of the corresponding gauge functional can result in different predictions of the string tension. By using non-trivial center regions, that is, regions whose boundary evaluates to a non-trivial center element, a resolution of this issue seems possible. We use such non-trivial center regions to guide simulated annealing procedures, preventing an underestimation of the string tension in order to resolve the Gribov copy problem.
\end{abstract}
\vspace{2cm}

\section{Introduction}
First proposed by 't Hooft \cite{THOOFT} and Cornwall  \cite{CORNWALL} the \textit{ center vortex model} gives an explanation of confinement in non-abelian gauge theories. It states that the vacuum is a condensate of quantised magnetic flux tubes, the so called \textit{vortices}. The vortex model is able to explain:
\begin{itemize}\setlength\itemsep{0mm}
\item behaviour of Wilson loops, see~\cite{DelDebbio:1998luz},
\item finite temperature phase transition $\rightarrow$ Polyakov loops,
\item orders of phase transitions in SU(2) and SU(3),
\item Casimir scaling of heavy-quark potential, see~\cite{Faber:1997rp},
\item spontaneous breaking of scale invariance, see~\cite{Langfeld:1997jx},
\item chiral symmetry breaking, see~\cite{Hollwieser, Faber:2017alm} $\rightarrow$ quark condensate,
\end{itemize}
but suffers from Gribov copy problems: predictions concerning the string tension depend on the specific implementation of the gauge fixing procedure, see \cite{Bornyakov:2002sb, Faber:2001hq}. In this work, an explanation of the problem is given before an improvement of the vortex detection is presented.

\begin{figure}[h!]
\begin{center}
\includegraphics[width=8.5cm]{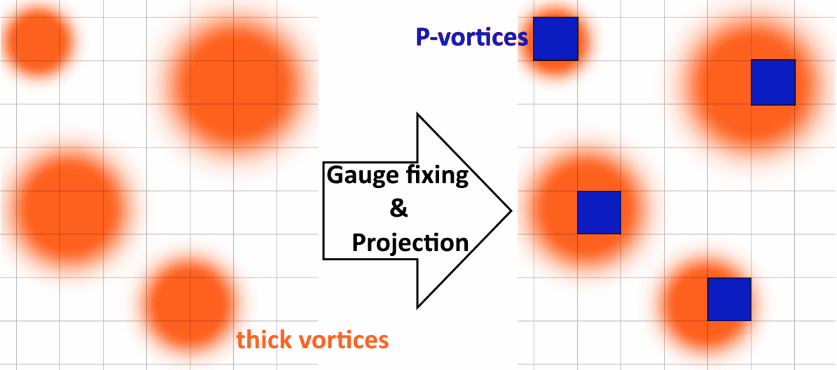}\end{center}
\caption{\label{BestFit} Vortex detection as a best fit procedure of P-Vortices to thick vortices shown in a two dimensional slice through a four dimensional lattice.}
\end{figure}

\newpage \noindent Center vortices are located by P-vortices, which are identified in direct maximal center gauge, the gauge which maximizes the functional
\begin{equation}
R^{2}= \sum_x \sum_\mu \mid \text{Tr}[ U_{\mu}(x)] \mid^2.
\label{FunctionalR}
\end{equation} 
The projection onto the center degrees of freedom
\begin{equation}
Z_{\mu}(x) = \text{sign } \text{Tr} [ U_{\mu}(x) ].
\label{Projection}
\end{equation} 
leads to plaquettes with non-trivial center values, P-plaquettes which form P-vortices, closed surfaces in dual space. This procedure can be seen as a best fit procedure of a thin vortex configuration to a given field configuration \cite{Faber, DelDebbio:1998luz}, see Figure \ref{BestFit}.
That P-vortices locate thick vortices is called \textit{vortex finding property}. 

\begin{figure}[h!]
\begin{center}
\includegraphics[width=7cm]{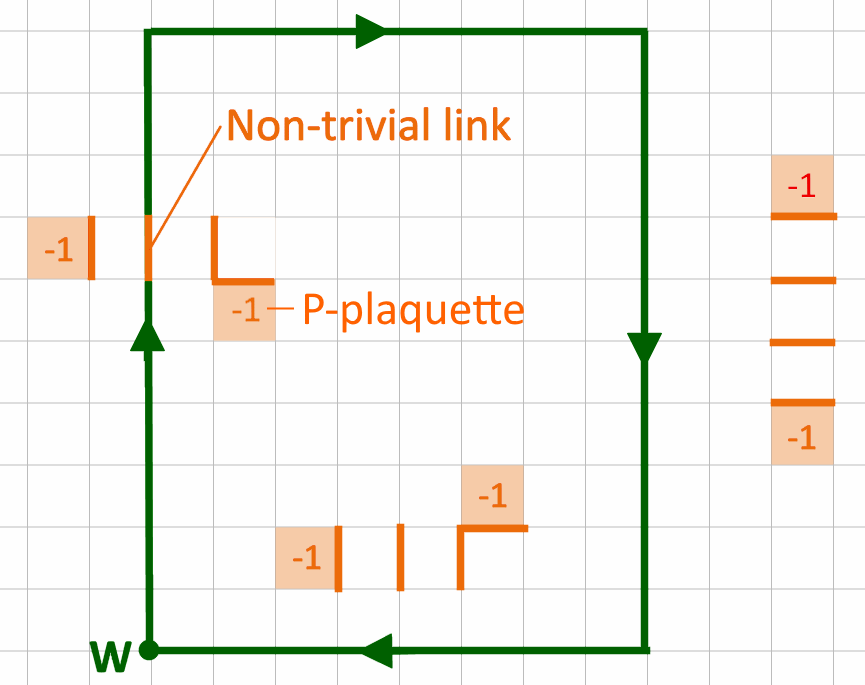}
\end{center}
\caption{\label{Piercing} Each P-plaquette contributes a non-trivial center element to surrounding Wilson loops.}
\end{figure}

Center vortices can be directly related to the string tension: the flux building up the vortex contributes a non-trivial center element to surrounding Wilson loops, see Figure \ref{Piercing}.
The behaviour of Wilson loops can be explained and a non vanishing string tension extracted by using the density $\rho_\mathrm{u}$ of uncorrelated P-plaquettes per unit volume
\begin{equation}
\begin{split}
\langle  \frac{1}{2} Tr(\boldsymbol{W}(R,T) \rangle = [\-1\; \rho_\mathrm{u} + 1 \; (1-\rho_\mathrm{u})]^{R \times T}  =  e^{\ln(1-2  \rho_\mathrm{u}) \; R \times T}
\Rightarrow \sigma = - \ln(1-2 \; \rho_\mathrm{u}).
\end{split}
\end{equation} 

\noindent The string tension can also be calculated by Creutz ratios
\begin{equation}
\chi(R,T)=-\ln\frac{\langle \boldsymbol{W}(R+1,T+1) \rangle \; \langle \boldsymbol{W}(R,T) \rangle}{\langle \boldsymbol{W}(R,T+1) \rangle \; \langle \boldsymbol{W}(R+1,T) \rangle}.
\end{equation}
From $ \langle \boldsymbol{W}(R,T) \rangle \approx e^{- \sigma \; R \; T - 2 \; \mu \; (R+T) + C} $ follows for sufficiently large R and T that $ \chi(R,T) \approx \sigma $. Creutz ratios for center projected Wilson loops are expected to give correct values for $\sigma$ if the vortex finding property is given.

\begin{figure}[h!]
\begin{center}
\includegraphics[width=4.5cm]{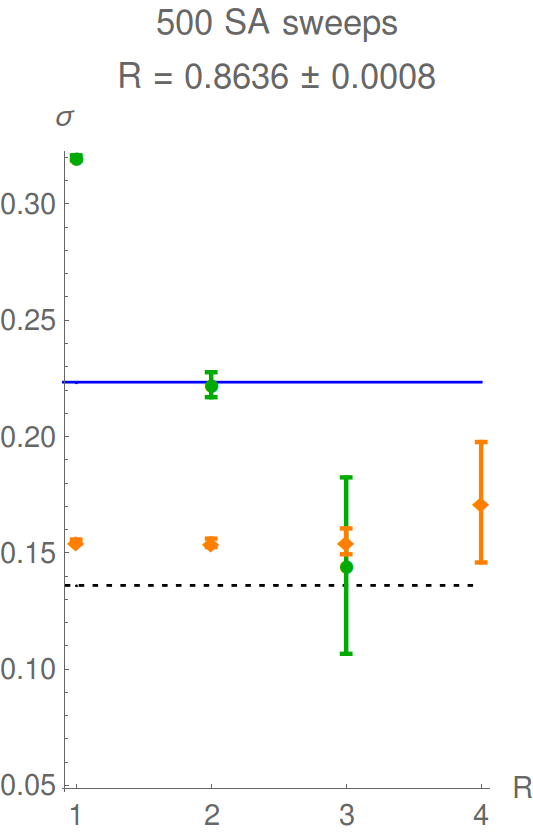}\hfill
\includegraphics[width=4.5cm]{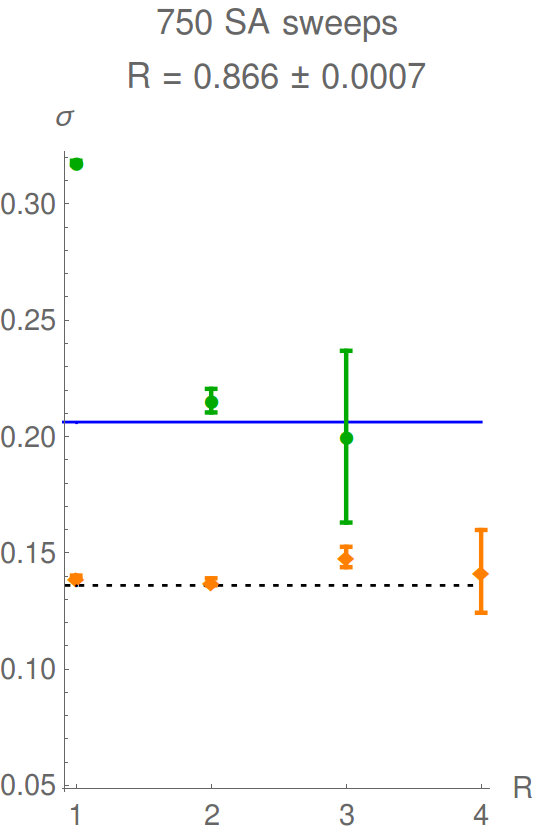}\hfill
\includegraphics[width=4.5cm]{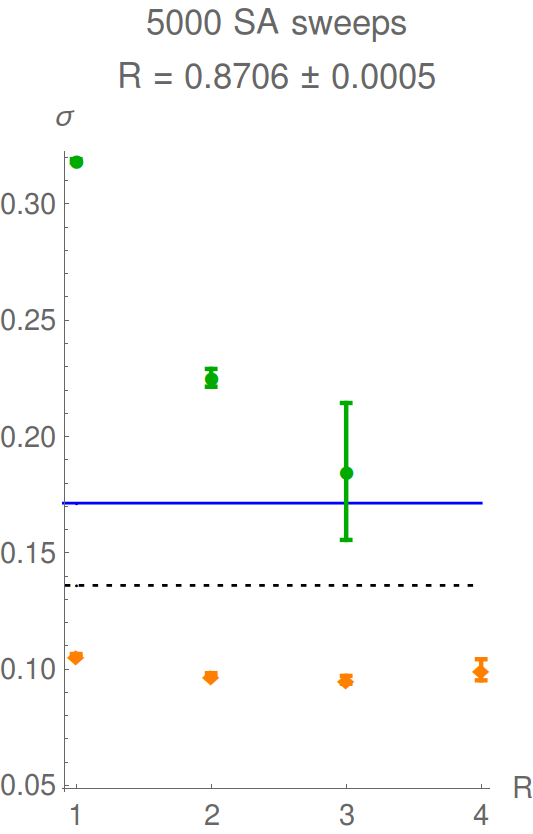}
\includegraphics[width=12cm]{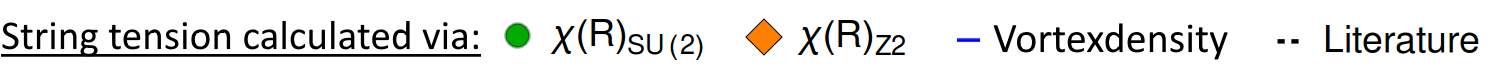}
\end{center}
\caption{\label{SigmaUnder} The string tension, calculated via Creutz ratios of the full theory $\chi(R)_{SU(2)}$, the center projected theory $\chi(R)_{Z2}$ and the vortex density. By increasing the number of simulated annealing sweeps a better value of gauge functional is reached, but the string tension underestimated by $\chi(R)_{Z2}$. The data was calculated in lattices of size $12^4$ (left),$12^4$ (middle) and $14^4$ (right) in Wilson action. The vortex density was not corrected for correlated P-plaquettes, hence is overestimated.}
\end{figure}

The problem with the direct maximal center gauge is that different local maxima of the gauge functional $R$ can lead to different predictions concerning the string tension in center projected configurations \cite{Bornyakov:2002sb, Faber:2001hq}. An improvement in the value of the gauge functional results in an underestimation of the string tension as can be seen in Figure \ref{SigmaUnder}. 
In fact, preliminary analysis show that the string tension decreases linearly with an improvement in the value of the gauge functional.

\begin{figure}[h!]
\begin{center}
\parbox{5cm}{
\includegraphics[width=5cm]{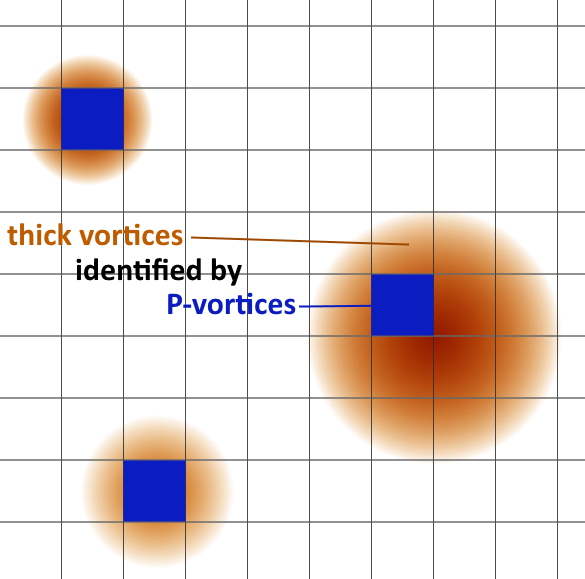}
}\parbox{2.1cm}{
\centering 
\Huge{$ \Longrightarrow $}
\\\small{loosing the\\vortex finding property}
}\parbox{5cm}{
\includegraphics[width=5cm]{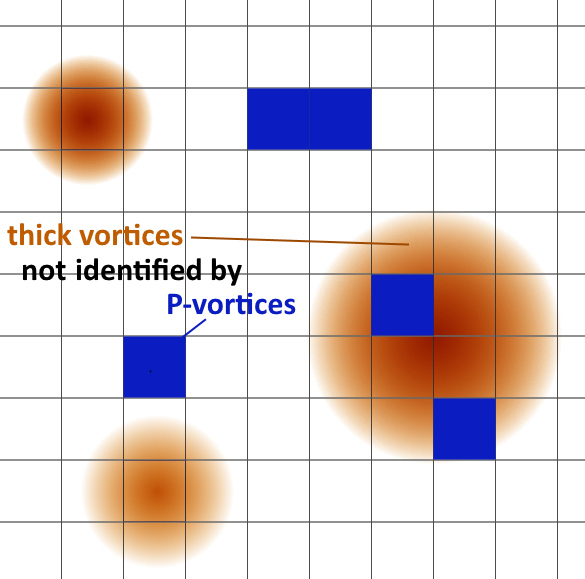}
}
\caption[Loss of the vortex finding property]{\label{LossVortexFindingProperty}When P-vortices no longer locate thick vortices, we speak of a loss of the \textit{vortex finding property}. The figure shows a two dimensional slice through a four dimensional lattice.}
\end{center}
\end{figure}

We believe that this is caused by a failing gauge fixing procedure during which the vortex finding property is lost. If the P-vortices fail to locate thick vortices the string tension will be underestimated by $\chi(R)_{Z2}$, see Figure \ref{LossVortexFindingProperty}.  
A failing vortex detection can result in vortex clusters disintegrating into small vortices consisting only of correlated P-plaquettes. This causes a misleadingly high vortex density.

The loss of the vortex finding property can be avoided by using the information about center regions, that is, regions, enclosed by a Wilson loop that evaluate to center elements.

Center regions can be related to a non-Abelian generalization of the Abelian stokes theorem: 
\begin{equation}
\begin{split}
 P \exp \left( i \oint_{\partial S} A_{\mu}(x) \; dx^{\mu} \right) = \mathcal{P} \exp \left( \frac{i}{2} \int_{S} \mathcal{F} _{\mu\nu}(x) \; dx^{\mu} \; dx^{\nu} \right), \qquad \qquad
\\[2mm]
\mathcal{F} _{\mu\nu}(x)= U^{-1}(x, O) \; F_{\mu\nu}(x) \; U(x,O), \qquad U(x,O) = P \exp \left( i \int_{l} A_{\eta}(y) \; dy^{\eta} \right),
\end{split}
\label{stokes}
\end{equation}
with $P$ denoting path ordering, $\mathcal{P}$ "surface ordering" and $l$ being a path from the base $O$ of $\partial S$ to $x$, see \cite{Broda:1995wv}. The left hand side of (\ref{stokes}) can be identified as the evaluation of an Wilson loop spanning the surface $S$.  The right hand side can be expressed using plaquettes: $U_{\mu\nu}(x)= \exp \left( i a^{2} F_{\mu\nu} + \mathcal{O}(a^{3})\right)$, with lattice spacing $a$, see \cite{GattringerLang}. With this ingredients the non-Abelian stokes theorem reads in the lattice as is shown in Figure \ref{stokes}:
\begin{figure}[h!]
\begin{center}
\parbox{2.0cm}{\centering \includegraphics[width=1.0cm]{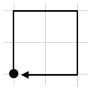}}\parbox{0.25cm}{$=$}
\parbox{2.0cm}{\centering \includegraphics[width=1.0cm]{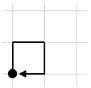}}\parbox{0.25cm}{$\times$}
\parbox{2.0cm}{\centering \includegraphics[width=1.0cm]{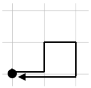}}\parbox{0.25cm}{$\times$}
\parbox{2.0cm}{\centering \includegraphics[width=1.0cm]{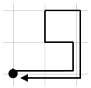}}\parbox{0.25cm}{$\times$}
\parbox{2.0cm}{\centering \includegraphics[width=1.0cm]{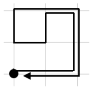}}
\caption[non-Abelian stokes theorem]{\label{CrAlgoGrowing}Factoring a Wilson into factors of plaquettes using the non-Abelian stokes theorem. }
\end{center}
\end{figure}

\begin{figure}[h!]
\begin{center}
\parbox{82mm}{
\parbox[c]{30mm}{\includegraphics[scale=0.7]{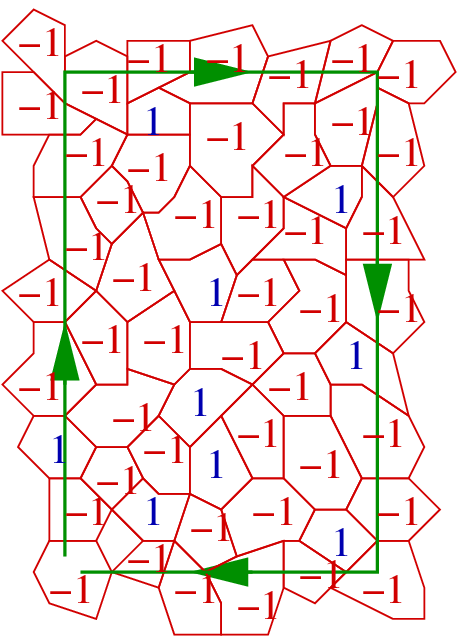}}
\parbox[c]{25mm}{$ \; = \;  (-1)^{22} \; \times$}
\parbox[c]{25mm}{\includegraphics[scale=0.7]{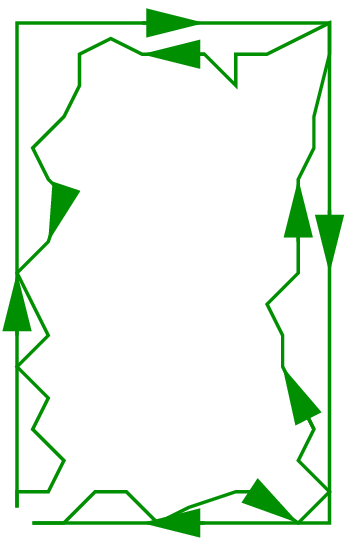}}
\\
\parbox{80mm}{
$ \; \underbrace{\qquad \qquad \qquad \qquad \quad}_{\text{\small center regions}} \quad \underbrace{ \quad \quad \quad}_{\text{\small Area law}} \quad \quad \underbrace{\qquad \qquad \qquad \quad}_{\text{ \small Perimeter law}}$}
} \hspace{2mm}\parbox[c]{6cm}{\small{\vspace{1mm}
Regions, whose boundaries evaluate to center elements can be used to factorize a Wilson loop into two parts: \begin{itemize}
\item \underline{an area factor} collecting the fully enclosed non-trivial regions, leading to a linear rising potential,\vspace{2mm}
\item \underline{a perimeter factor} from non-center contributions due to partially enclosed center regions.
\end{itemize}
}}
\end{center}
\vspace{-6.2mm}\caption[Factorization of a Wilson loop using center regions]{Center regions explain the coulombic behaviour and the linear rise of the quark anti-quark potential as they lead to an area law and a perimeter law for Wilson loops.}
\label{Factorization}
\end{figure}

By finding center regions, that is, plaquettes within $S$ that combine to bigger regions which evaluate to center elements, the Wilson loop spanning $S$ can be factorized into a commuting factor, a center element, and an non-Abelian part, see figure \ref{Factorization}. 

The center regions capture the center degrees of freedom and can be directly related to the behaviour of Wilson loops. It seems reasonable to demand that their evaluation should not be changed by center gauge or projection on the center degrees of freedom. We show that by preserving non-trivial center regions the loss of the vortex finding property is prevented and the full string tension can be recovered.

%%%%%%%%%%%%%%%%%%%%%%%%%%%%%%%%%%%%%%%%%%
%\newpage
\section{Materials and Methods}
The predictions of the center vortex model concerning the string tension in SU(2) gluonic quantum chromodynamic are analysed by calculating the Creutz ratios after center projection in maximal center gauge. The gauge fixing procedure is based upon simulated annealing, maximizing the functional~(\ref{FunctionalR}), that is, bringing each link as close to an center element as possible. The simulated annealing algorithms are modified, so  that the evaluation of center regions is preserved during the procedure: transformations resulting in non-trivial center regions projecting onto the non-trivial center element are enforced and transformations resulting in non-trivial center regions projecting onto the trivial center element are prevented.

\begin{figure}[h!]
\begin{center}
1)~\parbox[c]{4cm}{
\includegraphics[width=2.5cm]{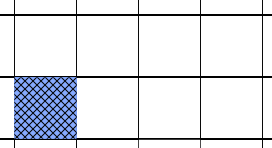}
} 2)~\parbox[c]{4cm}{
\includegraphics[width=2.5cm]{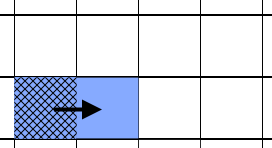}
} 3)~\parbox[c]{4cm}{
\includegraphics[width=2.5cm]{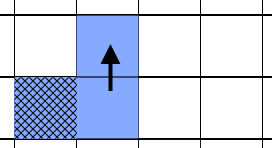}
}\\[1mm]
\parbox{14cm}{\footnotesize{\underline{Steps 1-3}: Starting with a plaquette that neither belongs to an already identified centre region, nor has already been taken as origin for growing a region, it is tested, which enlargement around a neighbouring plaquette brings the regions evaluation nearer to a centre element. Enlargement in best direction is done.}} 
\noindent 4)~\parbox[c]{4cm}{
\includegraphics[width=2.5cm]{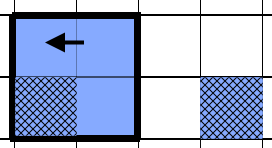}
} 5)~\parbox[c]{4cm}{
\includegraphics[width=2.5cm]{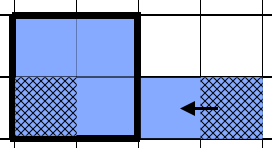}
} 6)~\parbox[c]{4cm}{
\includegraphics[width=2.5cm]{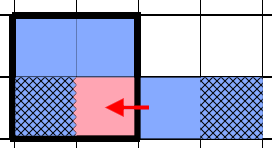}
}\\[1mm]
\parbox{14cm}{\footnotesize{\underline{Steps 4-6}: If no enlargement leads to further improvement, a new enlargement procedure is started with another plaquette. While its enlargement it can happen, that it would grow into an existing region. The collision handling described in following is used to prevent this:}}
\noindent 7a)~\parbox[c]{4.7cm}{
\includegraphics[width=2.5cm]{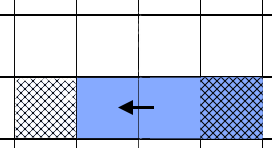}
} 7b)~\parbox[c]{4.7cm}{
\includegraphics[width=2.5cm]{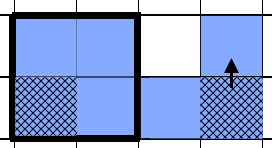}
}\\[1mm]
\parbox{6.5cm}{\footnotesize{\underline{Step 7a}: The evaluation of the growing region is nearer to a non-trivial centre element than the evaluation of the old region: Delete the old region, only keeping the mark on its starting plaquette and allow growing.}
}\hspace{0.5cm}\parbox{6.5cm}{\footnotesize{\underline{Step 7b}: The growing region evaluates further away from a non-trivial centre element than the existing one: prevent growing in this direction and, if possible, enlarge in second best direction instead. Multiple collisions after growing are possible.}}
\caption[Algorithm for region detection]{\label{CrAlgoOverall}The algorithm for detecting centre regions repeats these procedures until every plaquette either belongs to an identified region or has been taken once as starting plaquette for growing a region. The arrow marks the direction of enlargement. Plaquettes belonging to a region are coloured, plaquettes already used as origin are shaded.}\end{center}
\end{figure}

\pagebreak \noindent The detection of the non-trivial center regions of one lattice configuration is done by enlarging regions until their evaluation becomes nearest possible to an non-trivial center element, see Figure~\ref{CrAlgoOverall}. 
The algorithm starts with sorting the plaquettes of a given configuration by rising trace of their evaluation. This stack is worked down plaquette by plaquette, enlarging each as far as possible by adding neighbouring plaquettes. During this procedure, collisions of growing regions are prevented. 

\begin{figure}[h!]
\begin{center}
\includegraphics[width=12cm]{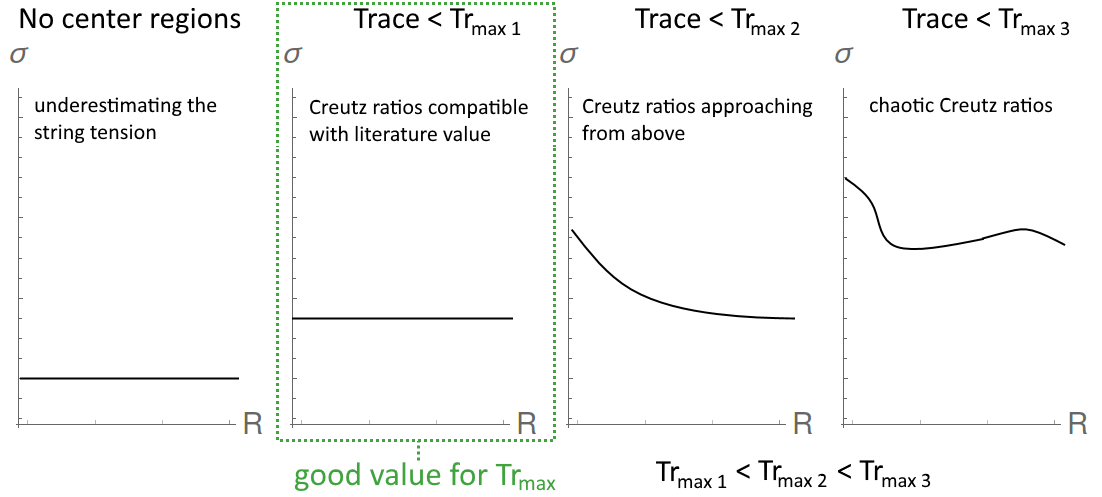}
\end{center}
\vspace{-6mm}\caption[Expected dependency of Creutz ratios on $Tr_\textrm{max}$]{$Tr_\textrm{max}$ can be fine tuned by looking at the dependency of the Creutz ratios on the loop size R.}
\label{CrBehaviour}
\end{figure}

The regions identified this way comprise also many, whose evaluation deviates far from the center of the group. A set of non-trivial center regions has to be selected from the set of identified regions: Only regions with traces smaller then $Tr_\textrm{max}$ are taken into account. This parameter $Tr_\textrm{max}$ has to be adjusted under consideration of the behaviour of Creutz ratios as shown in Figure \ref{CrBehaviour} which are calculated after gauge fixing and center projection. 

At low values of $Tr_\textrm{max}$ the Creutz ratios are expected to be nearly constant with respect to the loop size. With raising $Tr_\textrm{max}$ they start to approach their asymptotic value from above and become chaotic with $Tr_\textrm{max}$ chosen inappropriately high. 

As the center degrees of freedom are expected to capture the long range behaviour, the Creutz ratios calculated in center projected configurations are near to the correct value of the string tension already for small loop sizes. Hence we chose $Tr_\textrm{max}$ as high as possible without causing the behaviour of the Creutz ratios to approach the string tension from above.

The regions determined by this procedure are then used to guide the gauge fixing procedure. The influence on the predicted string tension is analysed by calculating the Creutz ratios in center projected configurations.

%%%%%%%%%%%%%%%%%%%%%%%%%%%%%%%%%%%%%%%%%%
\section{Results}
Here we present the calculations of the center vortex string tension for different values of $Tr_\textrm{max}$ at $\beta=2.3$. Similar results where obtained for $\beta=2.4$ and $\beta=2.5$. In the following, only the Creutz ratios of the center projected configurations $\chi(R)_{Z2}$ are of relevance. The Creutz ratios of the full SU(2) theory $\chi(R)_{SU(2)}$ and the calculations of the string tension based on the vortex density  are calculated for comparison. They are only shown for the sake of completeness. All data was calculated with SU(2) Wilson action.

\begin{figure}[h!]
\begin{center}
\includegraphics[height=6.5cm]{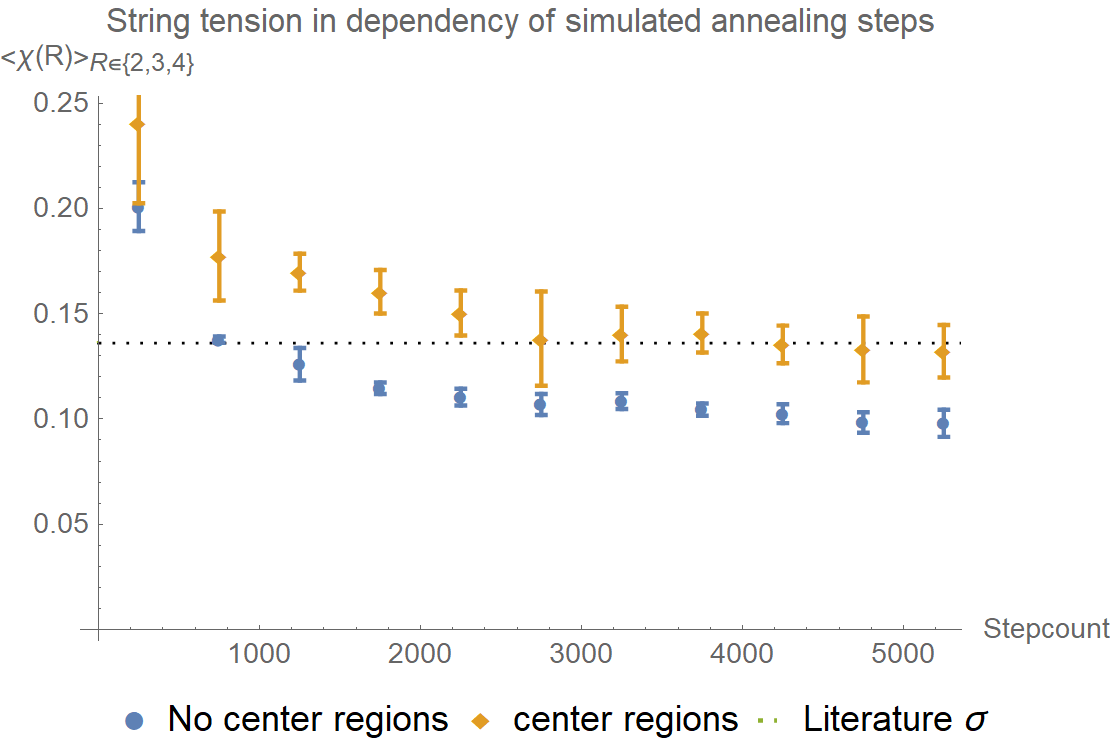}
\end{center}
\caption{By preserving center regions the Creutz ratios tend towards the literature value of string tension during the simulated annealing procedure. The data was calculated at $\beta=2.3$ in an $12^4$ lattice with 100 configurations taken into account per datapoint. Displayed is the mean of $\chi(2)$, $\chi(3)$ and $\chi(4)$. The increased error bars when center regions are preserved might be because the algorithm does not reach the exact local maxima, but fluctuates around it.}
\label{SA}
\end{figure}

\begin{figure}[h!]
\begin{center}
\includegraphics[height=6.8cm]{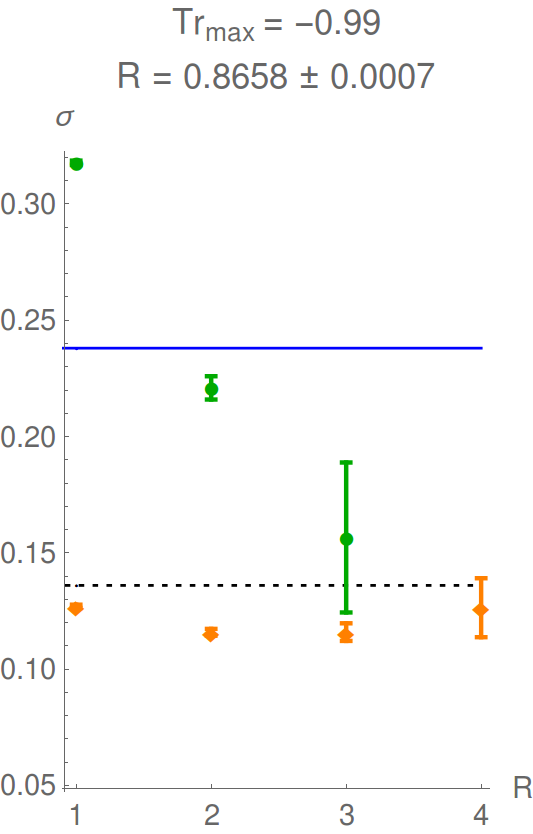}\hspace{6mm}
\vline \;
\includegraphics[height=6.8cm]{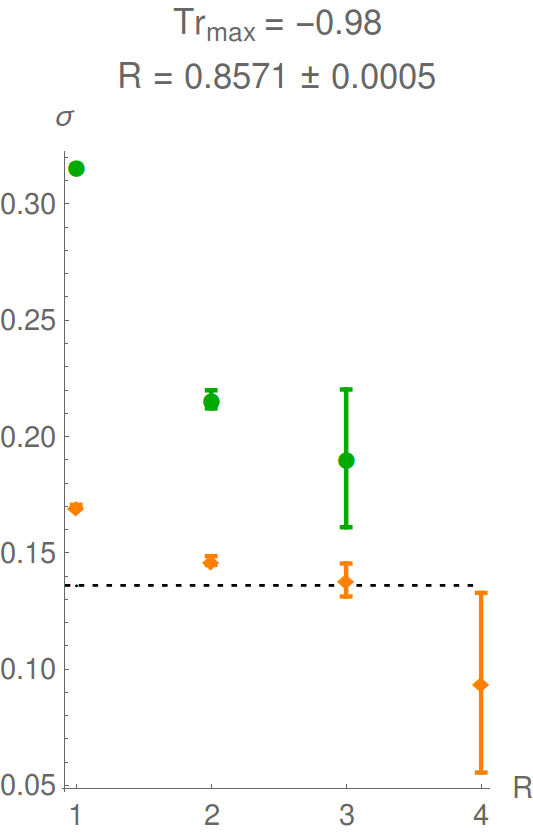}\hspace{6mm}
\includegraphics[height=6.8cm]{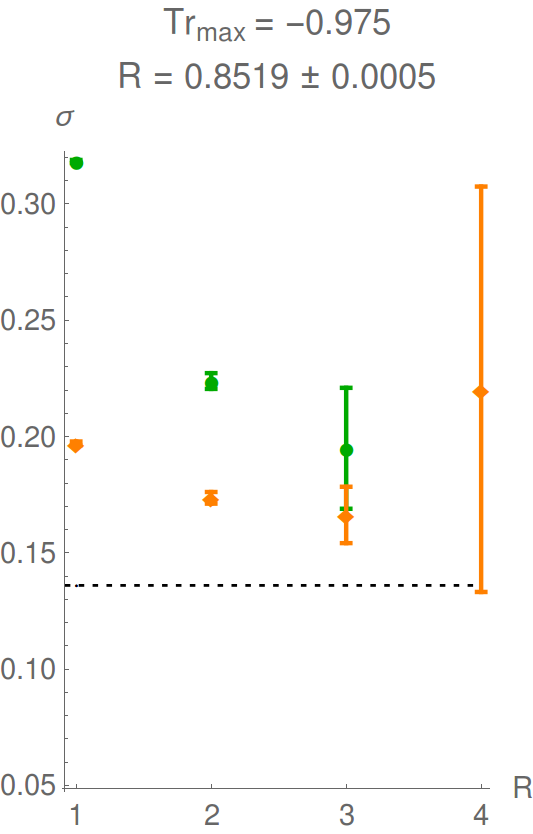}

$\overbrace{\hspace{3cm}}$\hspace{4.5cm}

\hspace{29mm}\parbox{5cm}{\includegraphics[height=7.1cm]{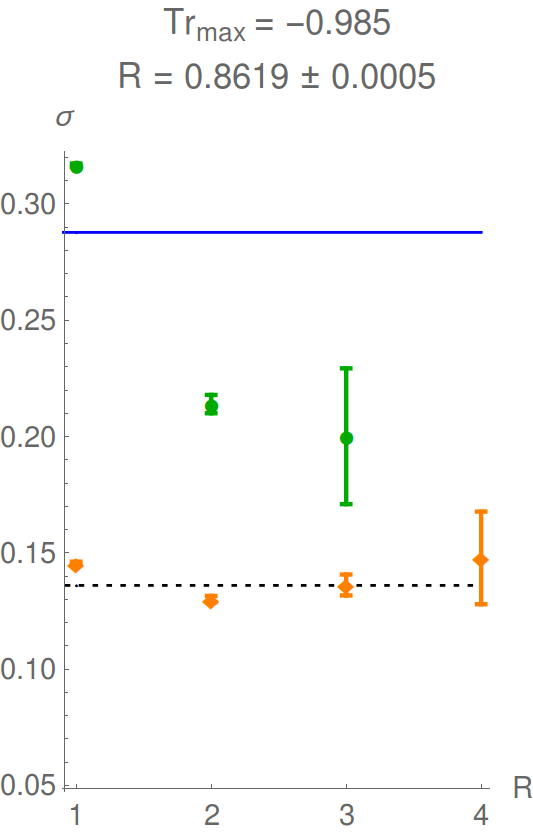}\hspace{1cm}}\textbf{versus}\hspace{1cm}\parbox{5cm}{
\includegraphics[height=7.1cm]{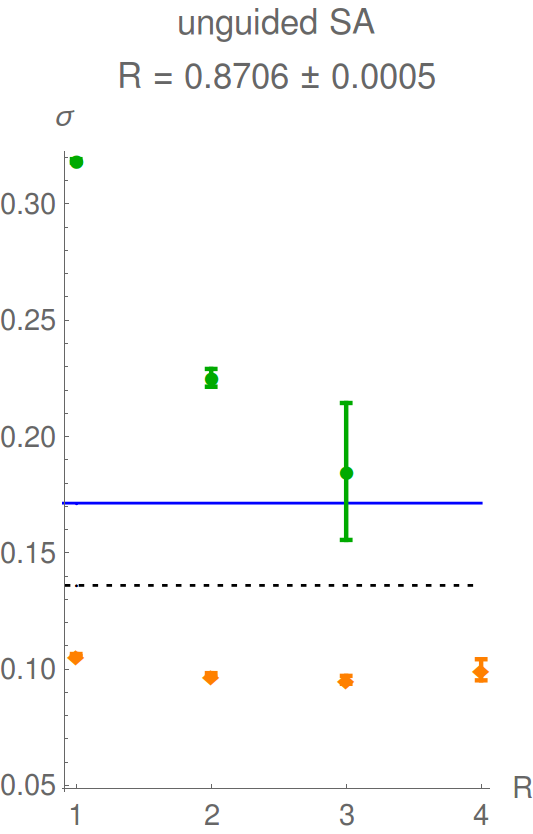}}

\includegraphics[width=12cm]{img/CreutzCr/legend.png}

\end{center}
\vspace{-6mm}\caption{Optimization of $Tr_\textrm{max}$ in the upper line and final results for the guided simulated annealing in the lower row at $\beta=2.3$. The creutz ratios where calculated with 300 Wilson configurations at $\beta=2.3$ in lattices of sizes $12^4$ in the upper left graph and $14^4$ for the other graphs. The error bars are calculated with the one-deletion-Jackknife method. The optimal value of $Tr_\textrm{max}$ was identified by taking into account the behaviour of the Creutz ratios and found to be around $Tr_\textrm{max}\approx -0.985$, reducing the value of gauge functional from $R=0.871$ to $R=0.862$. }
\label{ResultB2p3}
\end{figure}

The Creutz ratios tend towards the literature value of the string tension with increasing number of simulated annealing steps with a $Tr_\textrm{max}=-0.985$, whereas they clearly underestimate the string tension when center regions are ignored, see Figure \ref{SA}.
The full string tension can be easily recovered, although the value of the gauge functional is reduced, see Figure \ref{ResultB2p3}.
The upper three graphs show the calculations done for optimizing the value of $Tr_\textrm{max}$. The final results, shown in the left graph in the lower row are calculated with a value of $Tr_{nax}=-0.985$, that is, a value between the respective values of the left and middle graph in the upper line. The final results are compared with raw simulated annealing, that is, without preserving center regions shown in the right graph of the lower row. The large errors using center regions might result from fluctuations of the gauge functional around the maxima which can not be reached due to the constraint of the preservation of center regions: further approaches to the local maxima of the gauge functional are therefore prevented. 

%%%%%%%%%%%%%%%%%%%%%%%%%%%%%%%%%%%%%%%%%%
\pagebreak
\section{Discussion}
By preserving non-trivial center regions the full string tension can be recovered and extracted from the center degrees of freedom in SU(2) quantum chromodynamics. The choice of the free parameter $Tr_\mathrm{max}$ based on the behaviour of Creutz ratios does not give an unambiguous value, but merely an interval of good values of $Tr_\mathrm{max}$. This arbitrariness has to be investigated in further work. Preliminary data already hints at a way to eliminate it. The concept of identifying gauge independent observables evaluating to the relevant degrees of freedom and using them to guide the gauge fixing procedure reduces the number of free parameters of the gauge transformation. It forces all differing local maxima of the gauge functional to incorporate specific, gauge invariant properties that are related to the relevant degrees of freedom. This might be a solution to the Gribov copy problem wherever the gauge fixing procedure is based upon a specific gauge functional. The algorithms presented can be easily extended into higher symmetry groups or modified to capture different degrees of freedom. The procedures for identifying non-trivial center regions can also be used to reconstruct the thick vortices from P-plaquettes. This will allow further investigations of the color structure of vortices.
 
\paragraph*{Acknowledgement:} We thank to Vitaly Bornyakov for helpful discussions of the results.

\bibliography{qcdbib.bib}

\end{document}